\documentclass[a4paper,11pt,twocolumn]{article}

\usepackage{icphs2023}
\usepackage{metalogo} 
\usepackage{epstopdf}

\usepackage[safe]{tipa}

\usepackage[unicode=true,pdfusetitle,%
 pdfauthor={Pablo Perez Zarazaga, Zofia Malisz},%
 pdfkeywords={Whispered speech, Formant contours, Pitch contours, Intonation, Machine learning},%
 bookmarks=true,bookmarksnumbered=true,bookmarksopen=false,%
 breaklinks=true,pdfborder={0 0 0},backref=false,colorlinks=true,citecolor=blue]{hyperref}

\usepackage{xcolor,colortbl}
\usepackage{multirow}
\usepackage{tikz}
\usetikzlibrary{positioning}
\usetikzlibrary{arrows}
\usetikzlibrary{shapes}

\definecolor{Gray}{gray}{0.90}
\newcolumntype{b}{>{\columncolor{Gray}}c}
\newcolumntype{a}{>{\columncolor{white}}c}

\title{Recovering implicit pitch contours from formants in whispered speech}
\author{Pablo P\'erez Zarazaga, Zofia Malisz\thanks{This work was supported by the Swedish Research Council grant no.\ 2017-02861 ``Multimodal encoding of prosodic prominence in voiced and whispered speech''.}}
\organization{TMH, KTH Royal Institute of Technology, Stockholm}
\email{(\href{mailto:pablopz@kth.se}{pablopz}, \href{mailto:malisz@kth.se}{malisz})\href{mailto:pablopz@kth.se}{@kth.se}}
\begin{document}

\maketitle

\begin{abstract}
Whispered speech is characterised by a noise-like excitation that results in the lack of fundamental frequency. Considering that prosodic phenomena such as intonation are perceived through $f_0$ variation, the perception of whispered prosody is relatively difficult. At the same time, studies have shown that speakers do attempt to produce intonation when whispering and that prosodic variability is being transmitted, suggesting that intonation "survives" in whispered formant structure.

In this paper, we aim to estimate the way in which formant contours correlate with an "implicit" pitch contour in whisper, using a machine learning model. We propose a two-step method: using a parallel corpus, we first transform the whispered formants into their phonated equivalents using a denoising autoencoder. We then analyse the formant contours to predict phonated pitch contour variation. We observe that our method is effective in establishing a relationship between whispered and phonated formants and in uncovering implicit pitch contours in whisper.
\end{abstract}


\section{Introduction}
\vspace{-0.5em}
Pitch is an important feature in the encoding of prosody, and its variations over time are a defining characteristic of intonation. 
In whispered speech, fundamental frequency is not present in the signal as the glottis remains open and the vocal folds do not vibrate. In principle, this should prevent transmission of pitch and intonation via whisper. 

However, several studies~\cite{heeren2009perception,holmes1983acoustic,higashikawa1996perceived} have found that listeners perceive pitch and intonation effects in whispered speech. Similarly, it is evident that the whispered signal contains properties that offer cues to intonation. 

Formant raising, particularly of F1, is a well-attested characteristic of whispered speech relative to phonated speech ~\cite{zygis2017segmental, tran2010improvement, eklund1997comparative}. The higher formant frequency positions in whisper may be related to a more open configuration of the vocal tract similar to that found in e.g. Lombard speech~\cite{eklund1997comparative, zygis2017segmental}. In our previous work, we showed that the jaw is more open in whispered, relative to phonated, vowels in Swedish, supporting the hypothesis that whisper is a form of hyperspeech. However, others have suggested that formant raising is related to whispered pitch perception: ~\cite{higashikawa1996perceived} showed that listeners were able to discriminate between low and high-pitched whispered vowels on the basis of higher F1 and F2. Additionally, raised formant frequencies seem to connect to prosody-related laryngeal activity ~\cite{tran2010improvement}. ~\cite{coleman2002larynx} showed that in whispered speech production, laryngeal movements associated with pitch changes remain comparable to those in phonated speech. This indicates that prosody-related oral cavity shape modification takes place in whisper and hence is able to change the acoustics of the noise-like excitation travelling through the vocal tract - allowing for pitch inference in perception.
%

The harmonic relation between $f_0$ and formants has also been exploited in speech technology. In voice conversion systems, i.e. systems that reconstruct phonated speech from whispers, some solutions have tackled the problem of the missing $f_0$ by creating harmonic excitation for whispered speech using generative adversarial networks~\cite{shah2018novel,malaviya2020mspec}. \cite{mcloughlin2015reconstruction} showed promising results by adding an artificial pitch model to formant structure.

In the present paper, our process runs somewhat in the opposite direction: we first aim to learn from a) the relationship between $f_0$ and formants in phonated speech and b) its connection to the spectral properties of whispered speech. To this end, we present a machine learning method that models the variation in $f_0$ present in phonated speech through the variation in whispered formant values. One goal is to use these relationships to uncover and understand the "implicit" pitch contour in whisper that allows for the perception of prosody in the absence of $f_0$. The other is to support technological applications such as voice conversion.



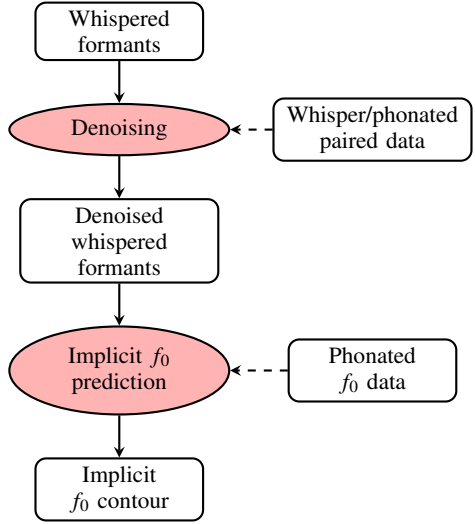
\begin{figure}[t]
\centering
\begin{tikzpicture}[auto, node distance = 7mm and 7mm,thick,scale=0.8, every node/.style={transform shape}]
        \node(wf) [draw, rectangle, rounded corners, text width=2.5cm, align = center, minimum height = 0.6cm] {Whispered formants};
        \node(Denoise) [draw, ellipse, fill=red!30!white, below =of wf, text width=2.3cm, align = center, minimum height = 0.6cm] {Denoising};
        \node(pf) [draw, rectangle, rounded corners, text width=3cm, align = center, minimum height = 0.6cm, below= of Denoise] {Denoised whispered formants};
        \node(Sequence) [draw, ellipse, fill=red!30!white, below =of pf, text width=2.3cm, align = center, minimum height = 0.6cm] {Implicit $f_0$ prediction};
        \node(f0c) [draw, rectangle, rounded corners, text width=2.5cm, align = center, minimum height = 0.6cm, below= of Sequence] {Implicit $f_0$ contour};

        \node(wpd) [draw, rectangle, rounded corners, text width=3cm, align = center, minimum height = 0.6cm, right= of Denoise] {Whisper/phonated paired data};
        \node(pd) [draw, rectangle, rounded corners, text width=2.5cm, align = center, minimum height = 0.6cm, right= of Sequence, xshift=0.25cm] {Phonated $f_0$ data};

        \draw[-stealth] (wf.south) -- (Denoise.north);
        \draw[-stealth] (Denoise.south) -- (pf.north);
        \draw[-stealth] (pf.south) -- (Sequence.north);
        \draw[-stealth] (Sequence.south) -- (f0c.north);
        \draw[dashed,-stealth] (wpd.west) -- (Denoise.east);
        \draw[dashed,-stealth] (pd.west) -- (Sequence.east);
\end{tikzpicture}
\caption{Steps of the proposed method. Red ellipses represent the implemented methods and white boxes represent the available data in each step. Dashed lines stand for the training process.}
\vspace{.5em}
\label{fig:pipeline}
\end{figure}

\section{METHOD}
\vspace{-0.5em}
\subsection{Data}
\vspace{-0.5em}
In this work, we use the CHAINS dataset~\cite{cummins2006chains}, which contains paired recordings in both phonated and whispered speech. CHAINS contains data from 20 male and 16 female speakers with Irish (28) and American (8) accents. The dataset consists of read sentences and text fragments in different speech modes such as natural phonated and whisper.  We use a subset of CHAINS with paired phonated and whispered sentences and analyse only vocalic phonemes.
%
%
%
 We transcribed the phonated audio samples using a Wav2Vec2 model~\cite{grosman2021wav2vec2} and copied the transcriptions into the whispered equivalents. The data was then labelled using the Montreal Forced Aligner~\cite{mcauliffe2017montreal} with a pre-trained model for American English. We observed that function word reductions produced event sequence mismatches between phonated and whispered pairs. They were subsequently removed from the analysis yielding an exactly matched dataset of 7549 phonated and whispered vowel phonemes. 


\begin{figure}
\centering
\includegraphics[width = .9\linewidth]{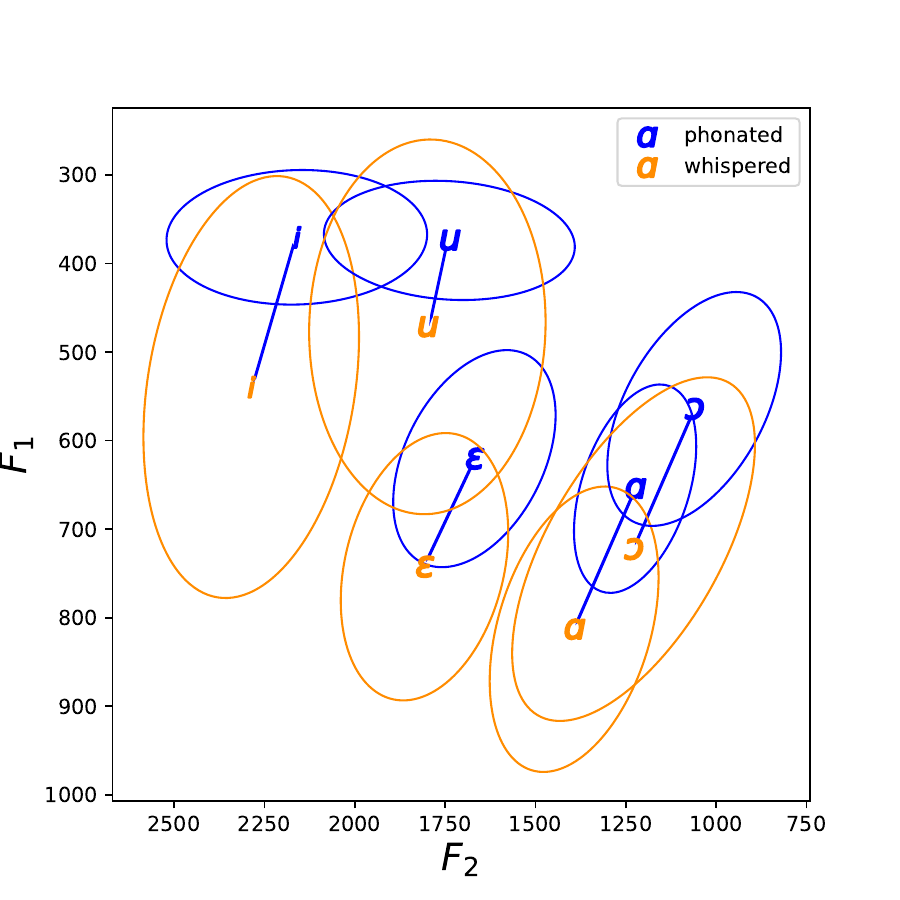}
\caption{Vowel space representing $F_1$ vs $F_2$ for the vowel sounds [\textipa{a}, \textepsilon, \textipa{i}, \textopeno, \textipa{u}] in phonated (blue) and whispered (red) speech and ellipses for their standard deviation.}
\label{fig:vowel}
\end{figure}

\subsection{Feature extraction and preprocessing}
\vspace{-0.5em}
We extracted $f_0$ and formant contours from vowel phones. $F_0$ values were extracted using Praat, with a time step of $0.01$~s and using a pitch floor and ceiling of 75 and 300~Hz respectively. To extract the formant contours, we used the Burg method in Praat with a time step of $0.01$~s, a $0.025$~s long analysis window and the maximum number of formants set to $5$. Pre-emphasis was applied over $50$~Hz. The formant ceilings were initialised to $5000$~Hz for male speakers and $5500$~Hz for female ones and optimised using the Escudero's method~\cite{escudero2009cross}. In order to normalise the length of each segment, we defined evenly distributed observation points every $10\%$ of the total duration of the phone. The contour measurements corresponding to $0\%$ and $100\%$ were then discarded to avoid instabilities due to formant transitions between phones, resulting in $9$ measurement points that define the pitch and formant contours. To reduce speaker variability in the formant contours, we normalised the data for each speaker following the Lobanov's normalisation method~\cite{lobanov1971classification}, where the data presents zero mean and unit variance with respect to its corresponding speaker. Additionally, pitch contours are also normalised to zero mean and unit variance relative to each speaker.

\subsection{Whispered pitch prediction}
\vspace{-0.5em}
We propose a machine learning method to estimate the "implicit" $f_0$ contour in whisper from whispered formant structure. The two-step method is depicted in Figure~\ref{fig:pipeline}. In the first step, we leverage the relationship between whispered and phonated formants residing in paired phones to provide a de-noised representation of the whispered formant frequencies. In the second step, we predict implicit, whispered $f_0$ contours by modelling the corresponding phonated $f_0$ with the denoised formant representations.   

\subsection{Formant denoising}
\vspace{-0.5em}
We assume that whispered formant structure is strongly correlated to its phonated equivalent. With the difference that in whisper, the formant frequencies are raised and the spectral envelope peaks are flatter manifesting a noisier behaviour. The vowel spaces obtained from the current dataset are presented in Figure~\ref{fig:vowel}. A high variance in some $F_1$ values can be observed due to the scarce representation of some phones in the dataset. The effects of whispered speech are reflected as an increase in the values of $F_1$ and $F_2$ and a higher variance in whispered, relative to phonated, formant values.

Hence, we consider whispered formant contours as a noisy representation of phonated formants and propose a denoising strategy to "transform" whispered formant contours to phonated equivalents. For this purpose, we use an autoencoder~\cite{lu2013speech} that learns higher-order statistical information from the formant contours. Using whispered-phonated formant pairs as input and output of the model, we aim to map the contours from whispered speech to the formant contours of the same phones in phonated speech.

The machine leaning method consists of an symmetric encoder-decoder structure. Three 1-D convolutional layers are used as filters over the time-dependent features ($16$ channels, kernel length of $3$). These layers, plus a fully-connected layer, generate encoder embeddings ($18$-sample) from whispered formant contours. The decoder uses three transposed convolutional layers to invert these embeddings into phonated speech formants.


We have observed that most of the contours in the data do not present significant variations. This might introduce bias in our network towards a constant value in the formant and pitch contours. In order to maximise the similarity with the target contour, we choose a different loss function: the cosine distance between target and predicted contours, such that:
\begin{equation}
\mathcal{L}_{cos} = - \frac{1}{N} \sum_{i = 0}^{N} \frac{\hat{y}_{i} \cdot y_{i}}{||\hat{y}_{i}||_{2}||y_{i}||_{2}}
\end{equation}

where $y_i$ and $\hat{y}_i$ stand for the target and denoised formant sequences of the $i_{th}$ sample and $\cdot$ represents the dot product between the two vectors.

The correlation coefficients for $F_1$, $F_2$ and $F_3$ are presented in Table~\ref{tab:corr_form}. We can observe that, while the correlation between phonated and denoised formant contours is improved for $F_1$, $F_2$, the network degrades the correlation for $F_3$ values. Therefore, we chose to apply denoised $F_1$ and $F_2$, while the contour of $F_3$ is used unmodified in the next step.



\subsection{Implicit $f_0$ prediction}
\vspace{-0.5em}
The second step in the method is a sequence prediction model that estimates the "implicit" $f_0$ contour based on the denoised contours of $F_1$, $F_2$ and (unmodified) $F_3$. This works under the assumption that the $f_0$ contour in the phonated data is related to the one "implied" in the whispered data. Admittedly, modelling the exact values of a non-existing pitch signal from formant contours is a challenging task. Therefore, this model focuses on estimating the relative changes in the pitch contour that represent intonation variability. 

We use a recurrent neural network (RNN), a deep learning method that is especially efficient in uncovering temporal dependencies within sequences. Considering the temporal dependencies in the formant and $f_0$ contours, a sequence-to-sequence model~\cite{sutskever2014sequence} should be particularly useful in modelling dependencies within and between dynamic contours. 
The input formant contours are processed by two bi-directional LSTM layers~\cite{schuster1997bidirectional} with $4$ hidden units, resulting in a sequence with $8$-dimensional features. The output sequence is then generated with a fully-connected layer that maps the corresponding $8$ features to the target pitch values.


Similarly to the formant denoising, we require a function that will maximise the similarity to the target sequence. The cosine distance can be seen as a normalised correlation, thus providing a better approximation to the sequence shape than mean squared error (MSE). Additionally, in order to also approximate the actual pitch values of the target, we will also include MSE in the loss function. The resulting loss function is then a combination of MSE and cosine distance:
\begin{equation}
\mathcal{L}_{seq} = \frac{1}{N}\sum_{i = 0}^{N} |\hat{y}_{i} - y_{i}|^{2} - \frac{1}{N} \sum_{i = 0}^{N} \frac{\hat{y}_{i} \cdot y_{i}}{||\hat{y}_{i}||_{2}||y_{i}||_{2}}.
\end{equation}

Both the denoising and prediction models were trained over $300$ epochs with an Adam optimiser and a learning rate of $10^{-4}$. Additionally, a recurrent dropout of 0.4 was added in the training loop to the LSTM layers of the $f_0$ prediction network.

\section{Results}
\vspace{-0.5em}

\begin{figure}
\vspace{-0.5em}
\centering
\includegraphics[width = \linewidth]{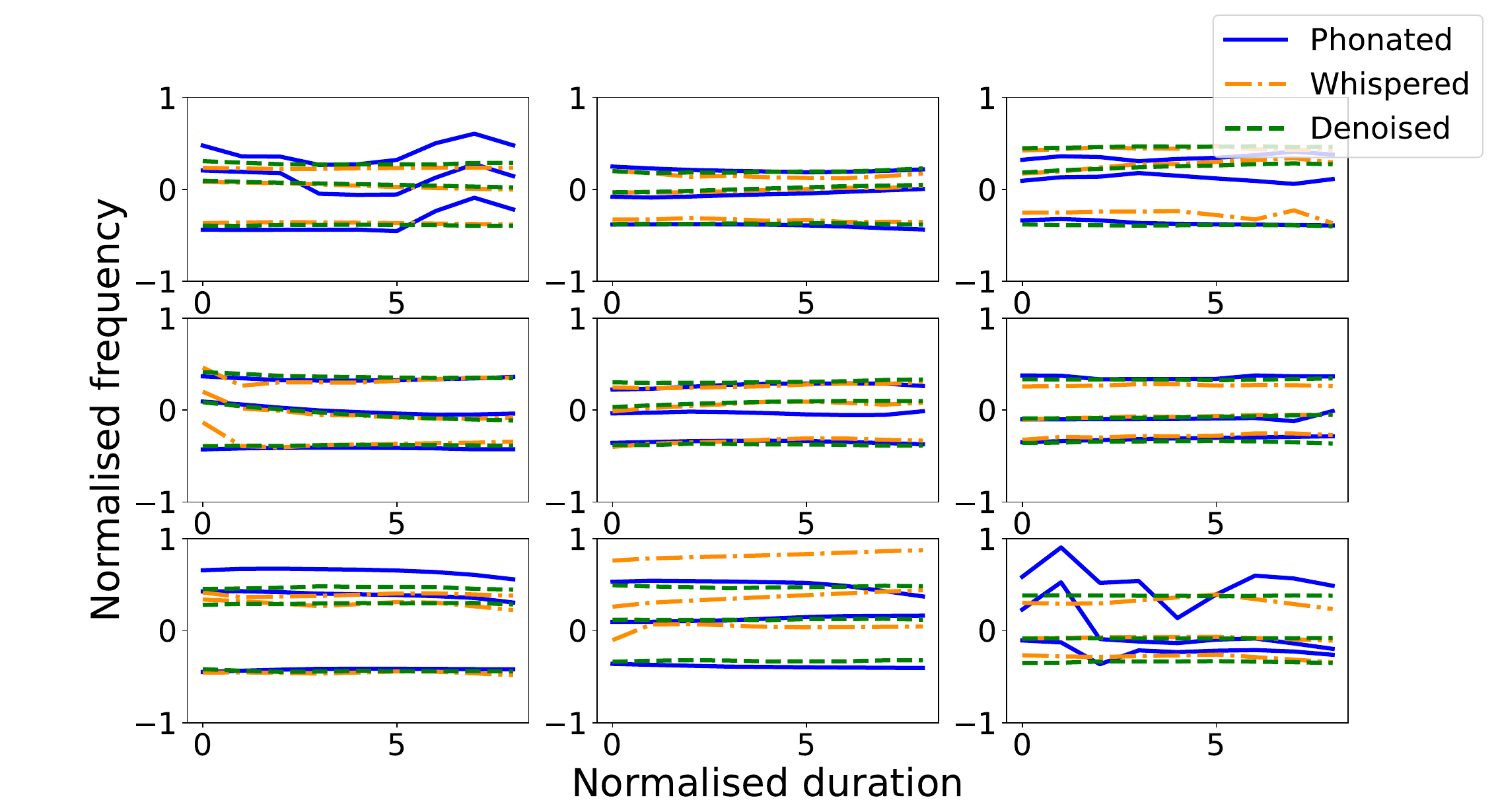}
\caption{Denoised formants from whispered contours with respect to original values.}
\label{fig:formant}
\vspace{1em}
\end{figure}

First of all, our results show that we were able to successfully leverage the close relationship between whispered and phonated speech to de-noise whispered formants. Table~\ref{tab:corr_form} shows that de-noised whispered formant contours exhibit a higher correlation with their phonated counterparts. The greatest improvement can be seen in $F_1$, known to exhibit the greatest difference between phonated and whispered speech~\cite{zygis2017segmental, tran2010improvement, eklund1997comparative}. In Figure~\ref{fig:formant}, we present several examples of the denoised formant contours in which the denoising model acts as a smoothing function. It is evident that the cosine similarity loss function allows us to closely estimate formant frequency contours in stationary vowel intervals found in the CHAINS corpus. Where the network fails is in some extreme cases with high contour variability, which may be due to various factors such as formant tracking errors, residual influence of the flanking phones etc. 

\begin{table}
\centering
\caption{Pearson's correlation coefficient of whispered and denoised formant contours with respect to phonated ones.}
\begin{tabular}{|c|c|c|}
\hline
Formant & Whispered & Denoised \\
\hline
$F_1$ & 0.19 & 0.34\\
\hline
$F_2$ & 0.44 & 0.47\\
\hline
$F_3$ & 0.26 & 0.16\\
\hline
\end{tabular}
\label{tab:corr_form}
\end{table}

\begin{figure}
\vspace{-0.5em}
\centering
\includegraphics[width = \linewidth]{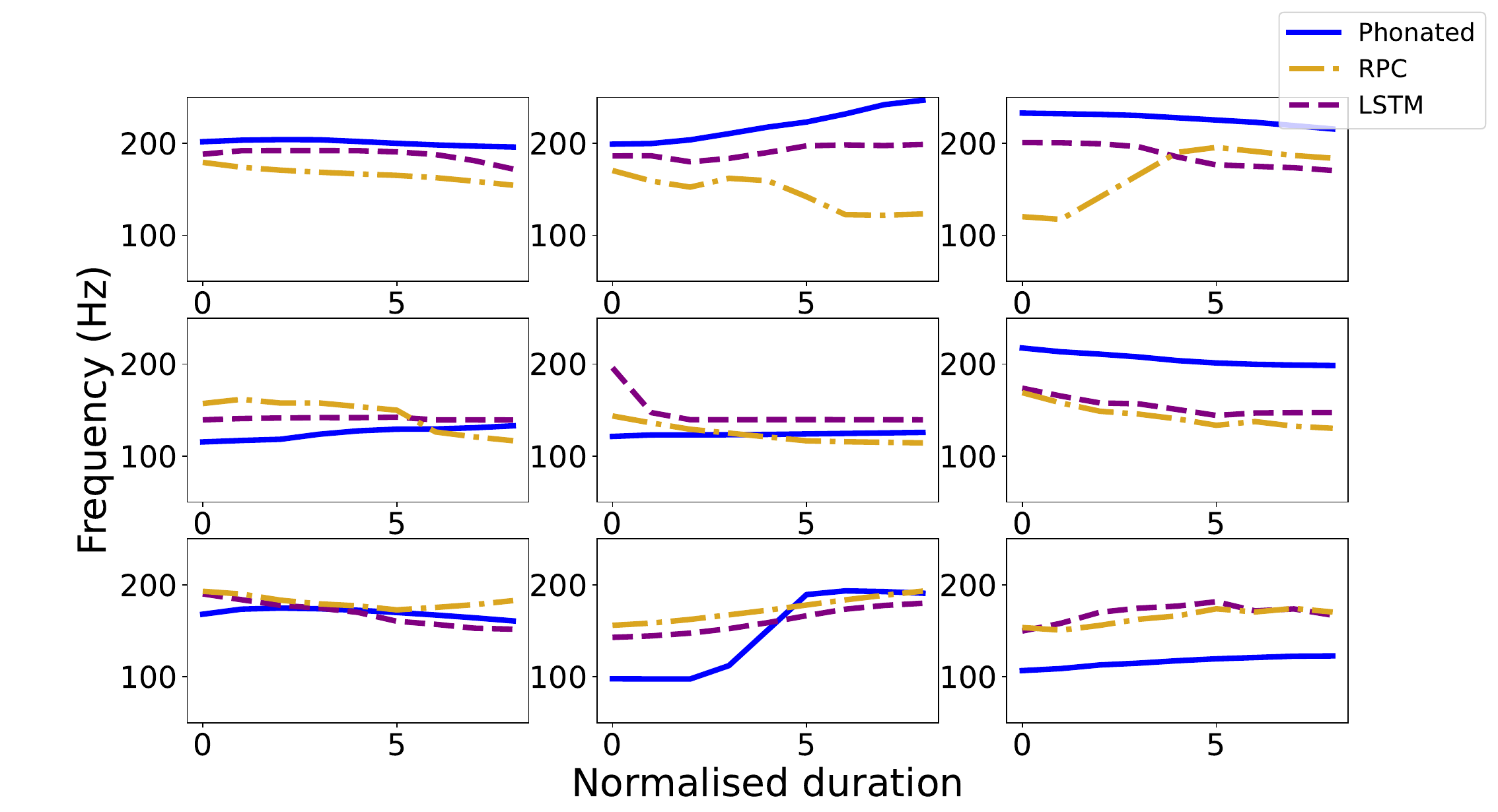}
\caption{Predicted implicit $f_0$ contour values compared to the original contours.}
\vspace{1em}
\label{fig:pitch}
\end{figure}

Second, the measured error and correlation values in the pitch prediction step are summarised in Table~\ref{tab:corr_pitch}. The refined pitch contour (RPC) from~\cite{mcloughlin2015reconstruction} is used as baseline. Our model obtains a positive correlation (Mean $r=0.56$, SD~$=~0.3$) between predicted and target contours that allows to generally follow the target pitch. Some examples of the resulting mapping between predicted and phonated contours can be observed in Figure~\ref{fig:pitch}. We see a correct direction of the trend in most cases. We also compare the absolute difference between the mean $f_0$ frequencies.  
Regarding comparisons to the RPC baseline, while the LSTM shows improvement in several individual cases, the average performance is similar to the baseline in correlation and error values. This leads us to believe that a more complex system such as LSTM would benefit from using additional input features supplying information beyond the minimalistic formant set.

\begin{table}[t]
\centering
\caption{Absolute error and correlation of implicit $f_0$ prediction from the refined pitch contour (RPC) estimation~\cite{mcloughlin2015reconstruction} and the proposed LSTM-based network with respect to phonated $f_0$.}
\setlength\tabcolsep{1.5pt}
\begin{tabular}{|c|c|c|c|c|}
\hline
\multirow{2}{*}{Method} & \multicolumn{2}{|c|}{RPC~\cite{mcloughlin2015reconstruction}} &  \multicolumn{2}{|c|}{LSTM} \\
 \cline{2-5} & Mean & St. Dev. & Mean & St. Dev.\\
\hline
Error (Hz) & 40 & 27 & 38 & 24\\
\hline
Correlation & 0.55 & 0.3 & 0.56 & 0.3\\
\hline
\end{tabular}
\label{tab:corr_pitch}
\end{table}

The absolute error from the predicted $f_0$ presents an average value of $38$ Hz, with a standard deviation of $24$ Hz. The number of different speakers (36) in this data was high and this error can be explained by the impact of speaker variability. This estimation could be improved by considering speaker-dependent features or tuning the model specifically for each speaker.

\section{Conclusion}
\vspace{-0.5em}
In this work, we have presented a machine learning method that estimates implicit $f_0$ contours from formant contours in whispered vowels. Our models have shown that it is possible to uncover an implicit $f_0$ trajectory in whisper, via its phonated equivalent, from $F_1$, $F_2$ and $F_3$ contours only.

In future lines of work, we will apply a speaker-dependent model. Our results show that the variability inherent in the 36 speakers modelled in this data has an influence on the $f_0$ prediction. Additionally, we would like to incorporate information on the analysed phones and their context into the formant analysis and pitch estimation. The present models showed promising results even before this variability was taken into account and we expect that adding the additional features will decidedly improve performance. 

This result will allow us to observe variations in perceived pitch and provide an improved analysis of intonation in whispered speech. The implementation of modern methods to uncover implicit pitch in whisper will also lead to developments in voice conversion systems.



\bibliographystyle{IEEEtran}
\bibliography{icphs2023}

\theendnotes

\end{document}